\documentclass[nohyper,12pt,letterpaper]{JHEP3}
\usepackage{graphicx,epsfig,youngtab}

\newcommand{\myfig}[3]{\begin{figure}[ht]
\begin{center}
\leavevmode \epsfxsize=#2cm \epsfbox{#1}
\end{center}
\caption{#3} \label{fig:#1}
\end{figure}}

\author{Robert de Mello Koch$^{1,2}$, Grant Mashile$^{1}$ and Nicholas Park$^{1}$\\
$^{1}$ National Institute for Theoretical Physics,\\
Department of Physics and Centre for Theoretical Physics,\\ 
University of the Witwatersrand,\\ 
Wits, 2050,\\ 
South Africa\\
\qquad\\
$^{2}$Stellenbosch Institute for Advanced Studies,\\
Stellenbosch,\\
South Africa\\
\qquad\\
E-mail: \email{robert@neo.phys.wits.ac.za, Grant.Mashile@students.wits.ac.za, Nicholas.Park@students.wits.ac.za}}

\abstract{In this article the anomalous dimension of a class of operators with a bare dimension of
$O(N)$ is studied. The operators considered are dual to excited states of a two giant graviton system. In the
Yang Mills theory they are described by restricted Schur polynomials, labeled with Young diagrams that
have at most two columns. In a certain limit the dilatation operator looks like a lattice version of a second
derivative, with the lattice emerging from the Young diagram itself.}

\preprint{WITS-CTP-052}

\title{Emergent Threebrane Lattices}


\def \Tr{\mbox{Tr\,}}
\def \threetwo{{}^3_2}
\def \twothree{{}^2_3}
\def \onetwo{{}^1_2}
\def \twoone{{}^2_1}

\def \onethree{{}^1_3}
\def \threeone{{}^3_1}

\begin{document}

\section{Introduction}

The AdS/CFT correspondence\cite{Maldacena:1997re} is a concrete realization of the conjectured
equivalence\cite{'tHooft:1973jz} between gauge theory and gravity, in 't Hooft's large $N$ limit. The 
identification of bulk strings with certain operators in the boundary theory was accomplished in 
\cite{Berenstein:2002jq}. The relevant operators take the form
$$ \Tr(YZZZYZZZZZYZZZZY) $$
A useful way of thinking about these operators, is that the $Z$s form a lattice on which the impurities 
($Y$) hop. Semiclassical states can be obtained by putting each lattice site in a coherent state. It is 
possible to derive a sigma model action which describes the relevant semiclassical dynamics. This action 
matches the low energy limit of the Polyakov action, providing a striking match between the string and 
Yang-Mills dynamics\cite{Kruczenski:2003gt,Hernandez:2004uw,Bellucci:2004qr}. 
What are the ingredients necessary for this matching?

\begin{itemize}

\item
As the ${\cal R}$-charge $J$ of the operators increase, they describe states in the string theory
of an increasing angular momentum. These states will expand as a consequence of the Myers effect.
The operators considered should contain $J=O(\sqrt{N})$ $Z$-fields and $O(1)$ impurities if they
are to expand to string sized objects.

\item
The dilatation operator closes on a set of single trace operators. To prevent mixing
with multitraces, it is necessary to take ${J^2\over N}\ll 1$. Notice that this is
consistent with taking $J=O(\sqrt{N})$.

\item
The dilatation operator acts (for example) on the second $Y$ in
$${\cal O}_l=\Tr (YZ^lYZ^{J-l})$$
to produce the combination $\propto {\cal O}_{l+1}+{\cal O}_{l-1}-2{\cal O}_{l}$. There are
two things worth noting. First, mixing between the operators ${\cal O}_l$ is highly constrained.
At large $N$ we have $\langle {\cal O}_l{\cal O}_k^\dagger\rangle\propto \delta_{kl}$. To one
loop, ${\cal O}_l$ can only mix with ${\cal O}_{l\pm 1}$. Second, the above linear combination 
clearly provides a lattice approximation to a second derivative. In this way the interpretation 
of the $Z$s as providing a lattice is natural and we see concretely how the string worldsheet 
emerges from the Yang-Mills theory.

\end{itemize}

This demonstration of stringy degrees of freedom in the Hilbert space of super Yang-Mills theory 
is encouraging, but is not a complete story. Indeed, since the discovery of 
D-branes\cite{Polchinski:1995mt} it has been clear that there is more to string theory than
strings - there are membrane excitations of various dimensionality in the theory. If the operator
in the Yang-Mills theory is to describe a threebrane of size $\approx 1$ in units of the radius
of curvature of the AdS space, we must consider operators with an ${\cal R}$-charge of $O(N)$.
These are the so-called giant gravitons\cite{McGreevy:2000cw}. For operators with such a large
${\cal R}$-charge there will be uncontrolled mixing between different trace structures, as a 
consequence of exploding combinatoric factors that over power the usual ${1\over N}$ suppression
of non-planar terms\cite{Balasubramanian:2001nh}. 
This difficulty was solved in \cite{Corley:2001zk} where it was shown that the Schur polynomials
have diagonal two point functions to all orders in ${1\over N}$. The Schur polynomials are labeled by
Young diagrams. 
The picture that naturally emerges\cite{Corley:2001zk} (see also
\cite{Berenstein:2004kk}) is that a Young diagram with $n$ long columns (a long column has
$O(N)$ boxes in it) is dual to a state of $n$ giant gravitons that have expanded in the S$^5$ of
AdS$_5\times$S$^5$; a Young diagram with $n$ long rows is dual to a state of $n$ giant gravitons 
that have expanded in the AdS$_5$ of AdS$_5\times$S$^5$.

Given that we know the operators dual to giant gravitons, it seems natural to
ask if we can compute the anomalous dimension of these operators. Further, by repeating
the argument that worked for strings, does the geometry of the threebrane\footnote{In this article the only threebrane
we consider is a giant graviton. Thus all threebranes have $S^3$ topology.} world volume emerge? 
Concretely, by acting with the dilatation operator on an operator dual to a giant graviton, does one 
see any hint of the giant worldvolume geometry?
There are a number of problems that need to be solved before this can be carried out:

\begin{itemize}

\item
One needs to include more than one matrix; all operators that are built out of a single complex Higgs
field preserve one half of the supersymmetries and hence are annihilated by the dilatation operator. 
The Schur polynomials are built out of $Z$s only. Fortunately, there is a multi-matrix generalization
of the Schur polynomials, the restricted Schur 
polynomials\cite{Balasubramanian:2004nb,de Mello Koch:2007uu,de Mello Koch:2007uv,Bekker:2007ea,Bhattacharyya:2008rb}.
The restricted Schur polynomial built out of $p$ matrices is labeled by $p+1$ Young diagrams.
Enhanced global non-abelian symmetries at zero coupling in Yang Mills theory provide a useful understanding (which generalizes
to other bases - see below)
of how the restricted Schur polynomials diagonalize the two-point functions of these multi-matrix operators\cite{Kimura:2008wy}.

\item
One would need to study more than just a single threebrane. Indeed, the small fluctuations
of a single threebrane\cite{Das:2000st,Sadri:2003mx} do not break supersymmetry, so consequently we expect any Schur polynomial
whose Young diagram labels have a single column or a single row (corresponding to a one threebrane state) will be
annihilated by the dilatation operator. We will explicitly demonstrate this.

\item
The results from\cite{Das:2000st} also suggest another difficulty:
you don't see radius dependence in the spectrum. This is not at all what
you would expect. Indeed, as the threebrane grows, you'd expect the wavelength
of the vibration modes to increase and consequently the energy of the mode 
to decrease. This naive expectation is not quite correct because the threebrane
expands in a nontrivial geometry. Due to the geometry, the modes of the larger
threebrane are blue shifted. This blue shifting exactly compensates the larger
wavelength so that the spectrum becomes independent of the size of the giant 
graviton.

\item
Taking the large $N$ limit usually provides a significant simplification, because
non-planar diagrams can be neglected. For the problem at hand it is not obvious
if there is any simplification.

\end{itemize}

We would like to study the simplest possible case of an excited giant graviton state.
It is clear that many features, including the emergence of the worldsheet lattice could be seen 
by considering the case of a BMN operator with just two impurities. With this motivation,
in this article we study operators with an ${\cal R}$-charge of order $N$ that contain
two impurities. Concretely, we built the operators using $n$ $Z$s and 2 $Y$s, with $n=O(N)$.
The operators are restricted Schur polynomials labeled by a Young diagram with $n+2$ boxes,
a Young diagram with $n$ boxes and a Young diagram with 2 boxes. Both the Young diagram with
$n+2$ boxes and the one with $n$ boxes have two large columns, so that we are studying a two 
threebrane state. The threebrane state carries two angular momenta. In the $Z$ plane it carries
$O(N)$ units of angular momentum whilst in the $Y$ plane it carries two units. 
When visualizing the two threebrane state, the angular
momentum in the $Y$ plane can be neglected so that the fluctuating threebranes will 
remain spherical, i.e. only the radius of the giant graviton can undergo fluctuations.
The giants will interact by means of open strings stretched between them. 
For this simple system, it is possible to anticipate the results. Each threebrane will
behave like a harmonic oscillator; the open strings stretched between these threebranes implies
that the oscillators are coupled. It is well known that coupled oscillators have two possible
normal modes, corresponding to the oscillators oscillating in phase or out of phase. We will
see that this physical picture does indeed emerge.

There are a number of papers discussing topics that are related to our study.
A related paper \cite{Berenstein:2003ah} with very similar goals, considers near maximal giant 
gravitons and their open string fluctuations at large $N$. 
For studies of strings attached to giant gravitons see
\cite{Balasubramanian:2004nb,Berenstein:2006qk,de Mello Koch:2007uu,de Mello Koch:2007uv,Bekker:2007ea}.
The basis provided by the restricted
Schur polynomials is only one of a number of possible bases. 
For other bases see \cite{Kimura:2007wy,Brown:2007xh,Brown:2008rr,Kimura:2008wy,Kimura:2009wy}.
The one loop dilatation operator in these bases has been studied in \cite{Brown:2008rr,Brown:2008rs}.
For a useful recent review of this material see \cite{Ramgoolam:2008yr}.
Finally, the papers \cite{tomyusuke} apply similar techniques to study operators of dimension
$O(\sqrt{N})$. These operators are dual to strings.

\section{Action of the Dilatation Operator}

We will consider the action of the one loop dilatation operator in the $SU(2)$ sector\cite{Beisert:2003tq}
$$ 
D = g_{\rm YM}^2 {\rm Tr}\,\big[ Y,Z\big]\big[ \partial_Y ,\partial_Z\big]
$$
on the restricted Schur polynomial 
$$
\chi_{(R,(r,s))}(Z^{\otimes \, n},Y^{\otimes \, 2})
=
{1\over n!2!}\sum_{\sigma\in S_{n+2}}{\rm Tr}_{(r,s)}(\Gamma_R(\sigma))
Z^{i_1}_{i_{\sigma(1)}}\cdots Z^{i_n}_{i_{\sigma(n)}}Y^{i_{n+1}}_{i_{\sigma(n+1)}}Y^{i_{n+2}}_{i_{\sigma(n+2)}}\, .
$$
The labels of our restricted Schur polynomial $\chi_{(R,(r,s))}$ are 
(i) $R$, which is a Young diagram with $n+2$ boxes or equivalently an irreducible representation of $S_{n+2}$,
(ii) $r$, which is a Young diagram with $n$ boxes or equivalently an irreducible representation of $S_n$ and
(iii) $s$ which is a Young diagram with $2$ boxes or equivalently an irreducible representation of $S_2$.
A simple calculation yields
\begin{eqnarray}
\label{basicreslt}
D \, \chi_{(R,(r,s))}(Z^{\otimes \, n},Y^{\otimes \, 2})=&&
{g_{\rm YM}^2\over (n-1)!}\sum_{\psi\in S_{n+2}}
\Tr_{(r,s)}\left(\Gamma_R ((n,n+2) \psi -\psi (n,n+2))\right)\times\cr
&&
\times Z^{i_1}_{i_{\psi (1)}}
\cdots
Z^{i_{n-1}}_{i_{\psi (n-1)}}
Y^{i_{n+1}}_{i_{\psi (n+1)}}
(YZ-ZY)_{i_{\psi (n)}}^{i_n}\delta^{i_{n+2}}_{i_{\psi (n+2)}}\, .
\end{eqnarray}
This can also be expressed as
$$
D \, \chi_{(R,(r,s))}(Z^{\otimes \, n},Y^{\otimes 2})=
g_{\rm YM}^2\Tr \left({d\over dV}\right) {1\over (n-1)!}\sum_{\psi\in S_{n+2}}
\Tr_{(r,s)}\left(\Gamma_R (\big[ (n,n+2), \psi\big] )\right)\times
$$
$$
\times 
Z^{i_1}_{i_{\psi (1)}}
\cdots
Z^{i_{n-1}}_{i_{\psi (n-1)}}
U_{i_{\psi (n)}}^{i_n}
Y^{i_{n+1}}_{i_{\psi (n+1)}}
V^{i_{n+2}}_{i_{\psi (n+2)}}\, ,
$$
where
$$
U=YZ-ZY\, .
$$
We would like to express this result as a sum over restricted
Schur polynomials. The basis that we are using is obtained by choosing an $S_n\times S_2$
subgroup of $S_{n+2}$. One way to construct the relevant representation of the subgroup, is to
remove two boxes from $R$ to obtain $r$; the two boxes which are removed can then be arranged
into irreducible representations of $S_2$. There is a second basis, which employs an $S_n\times S_1\times S_1$
basis. In this second basis, one simply keeps track of the order in which boxes were removed. The
removed boxes are not arranged into irreducible representations of $S_2$.

In the $S_n\times S_1\times S_1$ basis it is straight forward to evaluate the action of
$(n,n+2)$ in the $\Tr_{(r,s)}\left(\Gamma_R (\big[ (n,n+2), \psi\big] )\right)$ factor\cite{Bekker:2007ea}.
The action of the derivative in this basis has been worked out in  \cite{de Mello Koch:2004ws,de Mello Koch:2007uu}.
Finally, we need to ``separate'' the products of $X$ and $Z$ appearing in $U$. This can be done using the methods
developed in \cite{de Mello Koch:2007uv,Bekker:2007ea}. We will carry out this computation exactly, that is, to all
orders in ${1\over N}$.

The computation we performed can be summarized as
\begin{itemize}

\item{}Change from the $S_n\times S_2$ basis to the $S_n\times S_1\times S_1$ basis. The formulas for this
       change of basis are given in an Appendix.

\item{}Evaluate the action of $(n,n+2)$. The resulting character identities are given in an Appendix.

\item{}Perform the derivative with respect to $V$. In the $S_n\times S_1\times S_1$ basis, if $V$ is ``twisted''
       (in the notation of \cite{de Mello Koch:2007uu}) the derivative is zero. Thus we need only consider
       untwisted $V$ states. In this case, the matrix $V$ corresponds to a specific box in the Young diagram.
       The derivative removes this box and multiplies the result by the weight\footnote{Recall that the weight 
       of a box in row $i$ and column $j$ is $N-i+j$.} of the box; see
       \cite{de Mello Koch:2004ws,de Mello Koch:2007uu} for details.

\item{}Separate the products of $X$ and $Z$ appearing in $U$. This can be done using the methods
       developed in \cite{de Mello Koch:2007uv,Bekker:2007ea}. In \cite{de Mello Koch:2007uv,Bekker:2007ea}
       certain terms were dropped, since they were subleading at large $N$. In our calculation here we 
       derive new identities that allow us to treat these terms exactly. The identities and their derivation
       are given in an Appendix.

\item{}Change from the $S_n\times S_1\times S_1$ basis back to the $S_n\times S_2$ basis.

\end{itemize}

It is clear that the $S_n\times S_1\times S_1$ basis is more convenient for actually performing the computation.
Why not simply stick to this basis from the start? The $S_n\times S_2$ basis must be used, because it is only in 
this basis that we obtain a basis of operators whose free two point functions are orthogonal to all orders in
${1\over N}$. In fact, the $S_n\times S_1\times S_1$ basis is over complete; it would be the correct basis if the two
impurities were different fields.

\section{Single Threebrane States are Supersymmetric}

Single threebrane states are given by choosing $R$ to be a Young diagram with a single column (for a giant graviton
that has expanded in the S$^5$ of AdS$_5\times$S$^5$) or a single row (for a giant graviton that has expanded in
the AdS$_5$). These are both one dimensional representations of the symmetric group. Hence, the restrictions are
trivial and the representation is obviously Abelian. This implies that
$$
\Tr_{(r,s)}\left(\Gamma_R (\big[ (n,n+2), \psi\big] )\right) =
\Tr_{(r,s)}\left(\Gamma_R ((n,n+2))\Gamma_R( \psi) -\Gamma_R(\psi)\Gamma_R ((n,n+2))\right)=0
$$
so that (\ref{basicreslt}) vanishes. Thus, single threebrane states are supersymmetric. This is in perfect agreement
with the worldvolume analysis of \cite{Das:2000st,Sadri:2003mx}.

\section{Two threebranes and Two Impurities}

The general two threebrane states are given by the following operators
$$
O_a(b_0,b_1)=\chi_{\tiny \yng(2,2,2,1,1,1);\yng(2,2,2,1)\, \yng(1,1)},\qquad
O_b(b_0,b_1)=\chi_{\tiny \yng(2,2,2,1,1);\yng(2,1,1,1,1)\yng(1,1)}\, .
$$
$$
O_d(b_0,b_1)=\chi_{\tiny \yng(2,2,2,1,1);\yng(2,2,1,1)\yng(2)},\qquad
O_e(b_0,b_1)=\chi_{\tiny \yng(2,2,2,1,1);\yng(2,2,1,1)\yng(1,1)}\, .
$$
Once the representation of $S_2$ and the columns in $R$ from which these boxes are removed
has been given, we only need to specify the representation of $S_n$. This is given by specifying
the number of rows with two boxes in the row $=b_0$ and the number of rows with just a single
box in the row $=b_1$. We have reserved $O_c(b_0,0)$ for $O_d(b_0,b_1=0)$ because when $b_1=0$
there is no corresponding $O_e(b_0,b_1=0)$.

Acting with the dilatation operator on these two column restricted Schur polynomials, one 
generates terms corresponding to Schur polynomials that have an $R$ with three columns. The extra
operators are shown below 
$$
O_f(b_0,b_1)=\chi_{\tiny\yng(3,2,2,1,1,1,1);\yng(2,2,2,1,1,1)\yng(2)},\qquad
O_g(b_0,b_1)=\chi_{\tiny\yng(3,2,2,1,1,1,1);\yng(2,2,2,1,1,1)\yng(1,1)}\, .
$$
$$
O_h(b_0,b_1)=\chi_{\tiny\yng(3,2,2,2,1,1);\yng(2,2,2,1,1,1)\yng(2)},\qquad
O_i(b_0,b_1)=\chi_{\tiny\yng(3,2,2,2,1,1);\yng(2,2,2,1,1,1)\yng(1,1)}\, .
$$
In terms of these eight operators, the exact action of the dilatation operator is given in Appendix C.

The appearance of restricted Schur polynomials with three columns is a potential disaster. Indeed, the dilatation
operator acting on these three column restricted Schur polynomials will produce four column restricted 
Schur polynomials; acting on these will produce five column restricted 
Schur polynomials and so on. This would imply that we would have to consider the dilatation operator acting
on restricted Schur polynomials labeled by all possible Young diagrams $R$ with $n+2$ boxes - a much
more complicated problem. On physical grounds we would expect that at large $N$ the threebrane number would
be conserved - threebranes are stable semiclassical objects. 
Consequently, only two columns are long. Any extra columns are short. A restricted Schur polynomial
with two long columns and some additional short columns corresponds to a bound state of two threebranes plus
some KK gravitons. The transition from a two threebrane state to a state of two threebranes
plus KK gravitons involves graviton emission so this transition amplitude will be proportional to the string coupling.
In the 't Hooft limit, $g_s\propto {1\over N}$, so this transition will be suppressed. Consequently, we did not
expect mixing with three (or more) column restricted Schur polynomials.
Inspecting the result of Appendix C, its clear that the coefficients multiplying the three
column contributions are of the same size as the coefficients multiplying the two column contributions.
However, before we are able to decide whether terms are sub leading or not, we should write all of our expressions
in terms of operators normalized to have a unit two point function. The relevant two point functions
were computed using the results of \cite{de Mello Koch:2007uu,Bhattacharyya:2008rb} and are summarized in 
Appendix B. Notice that the three column terms are smaller by a factor of $b_0$ as compared to the two column
terms. This implies that the three column terms are suppressed by a factor of $1/\sqrt{b_0}$ and may therefore
be dropped if we take $b_0=O(N)$. 
This is very natural. Indeed, in the case of the BMN loops {\sl we restrict the ${\cal R}$-charge $J$ of
the loop to be $O(\sqrt{N})$ with $J^2/N\ll 1$ so that mixing with multi-trace states is suppressed.}
Here {\sl we restrict the ${\cal R}$-charge of the two threebrane state to be large enough that $b_0$
is $O(N)$ so that mixing with more than two column states is suppressed.}

We use hatted operators to denote normalized operators. The leading large $N$ action of the dilatation operator
on normalized operators is
{\small
$$ 
D\hat{O}_a (b_0,b_1)=
4g_{YM}^2{(N-b_0-b_1-1)\over (b_1+2)^2} \hat{O}_a(b_0,b_1)
$$
$$
-2g_{YM}^2{\sqrt{(N-b_0-b_1-1)(N-b_0+1)}\over b_1+2}\sqrt{b_1+3\over b_1+1}O_d(b_0,b_1)
$$
$$
+2g_{YM}^2{\sqrt{(N-b_0-b_1-1)(N-b_0+1)}b_1\over (b_1+2)^2}\sqrt{b_1+3\over b_1+1} \hat{O}_e(b_0,b_1)
$$
$$
+4g_{YM}^2{\sqrt{(N-b_0-b_1-1)(N-b_0+1)}\over (b_1+2)^2} \hat{O}_b(b_0-1,b_1+2)
$$
$$
+2g_{YM}^2{(N-b_0-b_1-1)\over (b_1+2)}\sqrt{b_1+1\over b_1+3} \hat{O}_d(b_0-1,b_1+2)
$$
$$
-2g_{YM}^2{(b_1+4)(N-b_0-b_1-1)\over (b_1+2)^2}\sqrt{b_1+1\over b_1+3} \hat{O}_e(b_0-1,b_1+2)
$$

$$ $$

$$
D \hat{O}_b(b_0,b_1) =
4g_{YM}^2{\sqrt{(N-b_0)(N-b_0-b_1)}\over b_1^2} \hat{O}_a(b_0+1,b_1-2)
$$
$$
-2g_{YM}^2{(N-b_0)\over b_1 }\sqrt{b_1+1\over b_1-1}\hat{O}_d(b_0+1,b_1-2)
$$
$$
+2g_{YM}^2{(N-b_0)(b_1-2)\over b_1^2}\sqrt{b_1+1\over b_1-1}\hat{O}_e(b_0+1,b_1-2)
$$
$$
+4g_{YM}^2{(N-b_0)\over b_1^2} \hat{O}_b(b_0,b_1)
$$
$$
+2g_{YM}^2{\sqrt{(N-b_0)(N-b_0-b_1)}\over b_1}\sqrt{b_1-1\over b_1+1}\hat{O}_d(b_0,b_1)
$$
$$
-2g_{YM}^2{(b_1+2)\sqrt{(N-b_0)(N-b_0-b_1)}\over b_1^2}\sqrt{b_1-1\over b_1+1}\hat{O}_e(b_0,b_1)
$$

$$ $$

$$
D\hat{O}_d(b_0,b_1)=
-2g_{YM}^2{\sqrt{(N-b_0+1)(N-b_0-b_1-1)}\over b_1+2}\sqrt{b_1+3\over b_1+1} \hat{O}_a(b_0,b_1)
$$
$$
+2g_{YM}^2{\sqrt{(N-b_0-b_1)(N-b_0)}\over b_1}\sqrt{b_1-1\over b_1+1}\hat{O}_b(b_0,b_1)
$$
$$
+g_{YM}^2\left( 2N-2b_0-b_1+3 \right)\hat{O}_d(b_0,b_1)
$$
$$
+g_{YM}^2{(N-b_0)(4-4b_1-2b_1^2)+b_1^3+b_1^2-4b_1\over b_1(b_1+2)} \hat{O}_e(b_0,b_1)
$$
$$
-2g_{YM}^2{(N-b_0+1)\over b_1+2}\sqrt{b_1+3\over b_1+1}\hat{O}_b(b_0-1,b_1+2)
$$
$$
-g_{YM}^2\sqrt{(N-b_0+1)(N-b_0-b_1-1)}\hat{O}_d(b_0-1,b_1+2)
$$
$$
+g_{YM}^2{\sqrt{(N-b_0+1)(N-b_0-b_1-1)}(b_1+4)\over (b_1+2)}\hat{O}_e(b_0-1,b_1+2)
$$
$$
+2g_{YM}^2{(N-b_0-b_1)\over b_1}\sqrt{b_1-1\over b_1+1}\hat{O}_a(b_0+1,b_1-2)
$$
$$
-g_{YM}^2\sqrt{(N-b_0)(N-b_0-b_1)}\hat{O}_d(b_0+1,b_1-2)
$$
$$
+g_{YM}^2{(b_1-2)\sqrt{(N-b_0)(N-b_0-b_1)}\over b_1}\hat{O}_e(b_0+1,b_1-2)
$$

$$ $$

$$
D \hat{O}_e(b_0,b_1)=
2g_{YM}^2{b_1\sqrt{(N-b_0+1)(N-b_0-b_1-1)}\over (b_1+2)^2}\sqrt{b_1+3\over b_1+1}\hat{O}_a(b_0,b_1)
$$
$$
-2g_{YM}^2{(b_1+2)\sqrt{(N-b_0-b_1)(N-b_0)}\over b_1^2}\sqrt{b_1-1\over b_1+1}\hat{O}_b(b_0,b_1)
$$
$$
+g_{YM}^2{(N-b_0)(4 -2b_1^2 -4b_1)+b_1^3+b_1^2-4b_1 \over b_1(b_1+2)}\hat{O}_d(b_0,b_1)
$$
$$
+g_{YM}^2{2(N-b_0)(b_1^4 + 4b_1^3 + 4b_1^2 - 8)-b_1^5-5b_1^4-8b_1^3+16b_1\over b_1^2(b_1+2)^2} \hat{O}_e(b_0,b_1)
$$
$$
+2g_{YM}^2{(N-b_0+1)b_1\over (b_1+2)^2}\sqrt{b_1+3\over b_1+1} \hat{O}_b(b_0-1,b_1+2)
$$
$$
+g_{YM}^2{\sqrt{(N-b_0+1)(N-b_0-b_1-1)}b_1\over (b_1+2)} \hat{O}_d(b_0-1,b_1+2)
$$
$$
-g_{YM}^2{\sqrt{(N-b_0+1)(N-b_0-b_1-1)}b_1(b_1+4)\over (b_1+2)^2} \hat{O}_e(b_0-1,b_1+2)
$$
$$
-2g_{YM}^2{(N-b_0-b_1)(b_1+2)\over b_1^2}\sqrt{b_1-1\over b_1+1} \hat{O}_a(b_0+1,b_1-2)
$$
$$
+g_{YM}^2{\sqrt{(N-b_0)(N-b_0-b_1)}(b_1+2)\over b_1} \hat{O}_d(b_0+1,b_1-2)
$$
$$
-g_{YM}^2{(b_1+2)(b_1-2)\sqrt{(N-b_0)(N-b_0-b_1)}\over b_1^2} \hat{O}_e(b_0+1,b_1-2)
$$
}
Notice that mixing is highly suppressed. Indeed, these operators are orthogonal, at tree level, to all order in
$1/N$. From the above expression we see that at one loop operators can only mix if they differ by at most,
by one box in their $R$ label. This suppression has been observed before in other
bases \cite{de Mello Koch:2007uv,Bekker:2007ea,Brown:2008rs}. Another point worth noting is that if we take the usual 't Hooft limit
$N\to\infty$ with $\lambda=g_{YM}^2 N$ fixed, the one loop anomalous dimension is $O(1)$. The usual 't Hooft
limit thus leads to a well defined and non-trivial problem.

\section{Emergence of the Radial Direction}

Recall that the ${\cal R}$-charge of an operator in the field theory maps into the angular momentum of the dual string
theory state and that the angular momentum of the string theory state determines its size. Identifying the two columns
with the two threebranes, the number of boxes in each column determines the angular momentum and hence the size of each
threebrane. In the limit that $N-b_0=O(N)$, $b_0=O(N)$ and $b_1=O(\sqrt{N})$ we have non-maximal giants which are separated 
by a distance of $O(1)$ in string units. In this limit, we expect the dynamics to simplify. The action of the dilatation
operator becomes

$$ 
D\hat{O}_a (b_0,b_1) = \lambda\times O\left({1\over b_1}\right)
$$
$$
D \hat{O}_b(b_0,b_1) = \lambda\times O\left({1\over b_1}\right)
$$

{\vskip 0.75cm}

$$
D\hat{O}_d(b_0,b_1)=
\lambda (1-{b_0\over N})\left(2\hat{O}_d(b_0,b_1)-\hat{O}_d(b_0-1,b_1+2)-\hat{O}_d(b_0+1,b_1-2)\right)
$$
$$
-\lambda (1-{b_0\over N})\left(2\hat{O}_e(b_0,b_1)-\hat{O}_e(b_0-1,b_1+2)-\hat{O}_e(b_0+1,b_1-2)\right)
+\lambda\times O\left({1\over b_1}\right)
$$

{\vskip 0.75cm}

$$
D\hat{O}_e(b_0,b_1)=
\lambda (1-{b_0\over N})\left(2\hat{O}_e(b_0,b_1)-\hat{O}_e(b_0-1,b_1+2)-\hat{O}_e(b_0+1,b_1-2)\right)
$$
$$
-\lambda (1-{b_0\over N})\left(2\hat{O}_d(b_0,b_1)-\hat{O}_d(b_0-1,b_1+2)-\hat{O}_d(b_0+1,b_1-2)\right)
+\lambda\times O\left({1\over b_1}\right)
$$

These results have a natural interpretation. It looks as if $\hat{O}_a(b_0,b_1)$, $\hat{O}_b(b_0,b_1)$
and $\hat{O}_d(b_0,b_1)+\hat{O}_e(b_0,b_1)$ remain supersymmetric. First, note that it is natural to 
interpret $\hat{O}_a(b_0,b_1)$ as a state in which we deform only the smaller threebrane. Recall that deforming a
single threebrane gives us a supersymmetric state so it seems natural for $\hat{O}_a(b_0,b_1)$ to remain supersymmetric.
Similarly, $\hat{O}_b(b_0,b_1)$ can be interpreted as a state in which we deform only the smaller threebrane and a 
similar comment can be made. The fact that the combination $\hat{O}_d(b_0,b_1)+\hat{O}_e(b_0,b_1)$ is annihilated
by $D$ suggests that there is also a supersymmetric way to deform the pair of threebranes. Finally, notice that
if we set $\hat{O}_d(b_0,b_1)-\hat{O}_e(b_0,b_1)\equiv \hat{O}_{d-e}(b_0,b_1)$ we have
$$
D\hat{O}_{d-e}(b_0,b_1)=
2\lambda (1-{b_0\over N})\left(2\hat{O}_{d-e}(b_0,b_1)-\hat{O}_{d-e}(b_0-1,b_1+2)-\hat{O}_{d-e}(b_0+1,b_1-2)\right)\, .
$$
The right hand side again looks like a discretization of the second derivative. This time {\sl it is the Young diagram
itself that is defining the lattice!} This result looks rather intuitive, especially after recalling that the number of 
boxes in each column sets the angular momentum and hence the radius of the corresponding threebrane. 

\section{Numerical Results}

We have not managed to analytically solve for the spectrum of the one loop anomalous dimension.
This would entail finding the eigenvalues and eigenvectors of the last equations given in section 4.
It is however straight forward to solve for the spectrum numerically. The spectrum is given in figure
1. 

\myfig{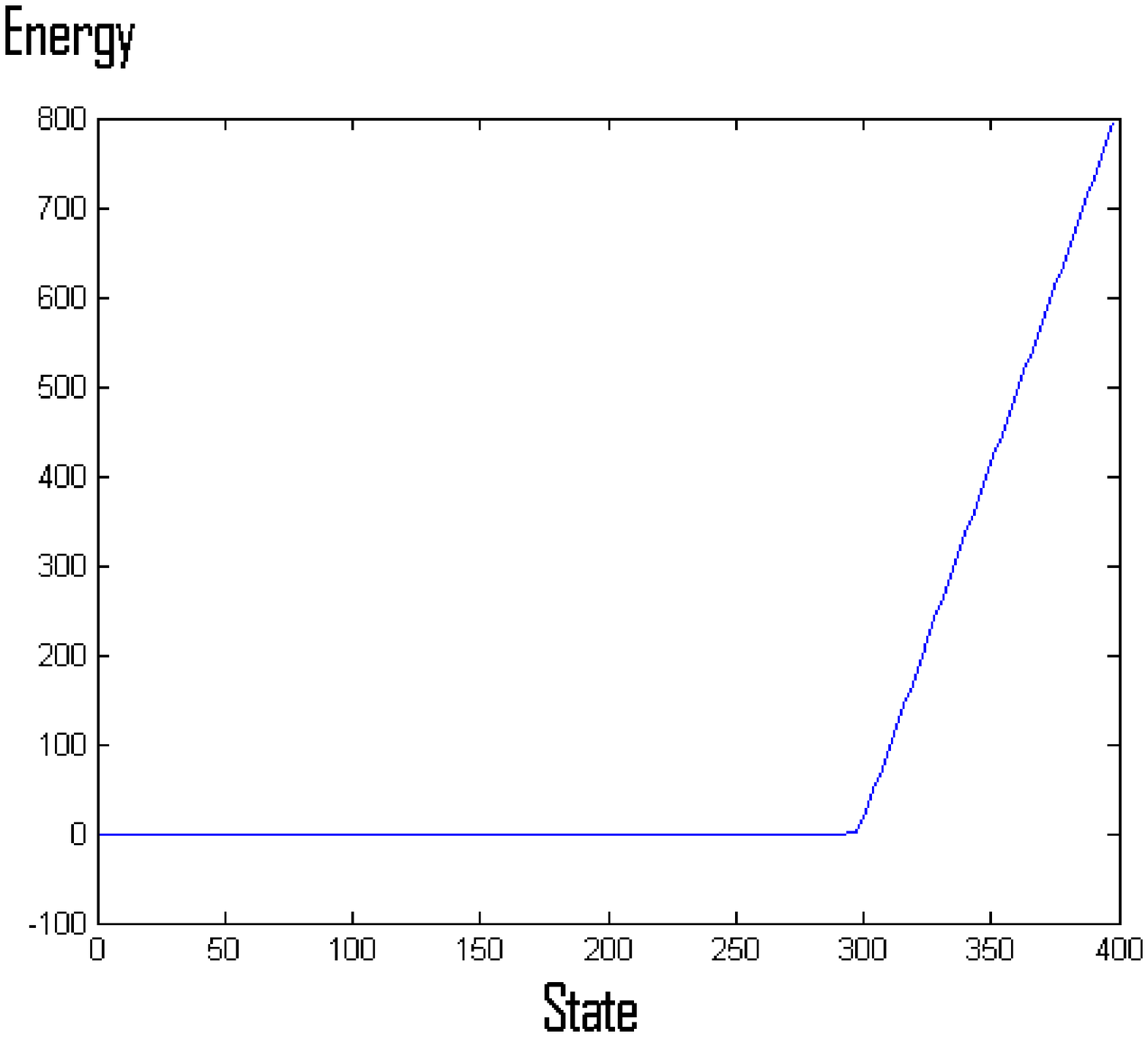}{10.0}{The spectrum of anomalous dimensions $=$ the energy of the dual giant graviton state.
To produce this plot we used $N=100000$ and considered an operator built from $199800$ $Z$s and 2 $Y$s. There
are a total of 398 excited two threebrane states.}

The example shown has $N=100000$ and considers operators built from $199800$ $Z$s and 2 $Y$s. There
are a total of 398 excited two threebrane states and 297 of them have zero energy. If the total number of 
states is $N_{\rm states}$ we find that $0.75(N_{\rm states}-2)$ are massless. The results of the last
section would have suggested that $3/4$ of the states are zero energy, so this is not unexpected.
What we did find rather remarkable, is that the states with non-zero energy have a constant energy spacing.
This strongly suggests that this system is secretly a harmonic oscillator; we have not however been able to
demonstrate this analytically.  
In the previous section we have seen that the equation for the non-zero energy states is
written in terms of a discretized version of the second derivative. To obtain the spectrum we need to know
what boundary conditions are to be applied. Notice that these boundary conditions are coded into the equations 
for the action of $D$. As $b_1$ gets smaller and smaller the factors of $b_1$ and $b_1-1$ prevent the first
column from shrinking to a size smaller than the second column. When the two columns are the same size, the
first column can't shrink any further - this is one boundary. As the first column grows we get to a point where
$b_0=N$. At this point, the factors of $N-b_0$ prevent the first column from growing to an even larger size. This
is the second boundary.


\myfig{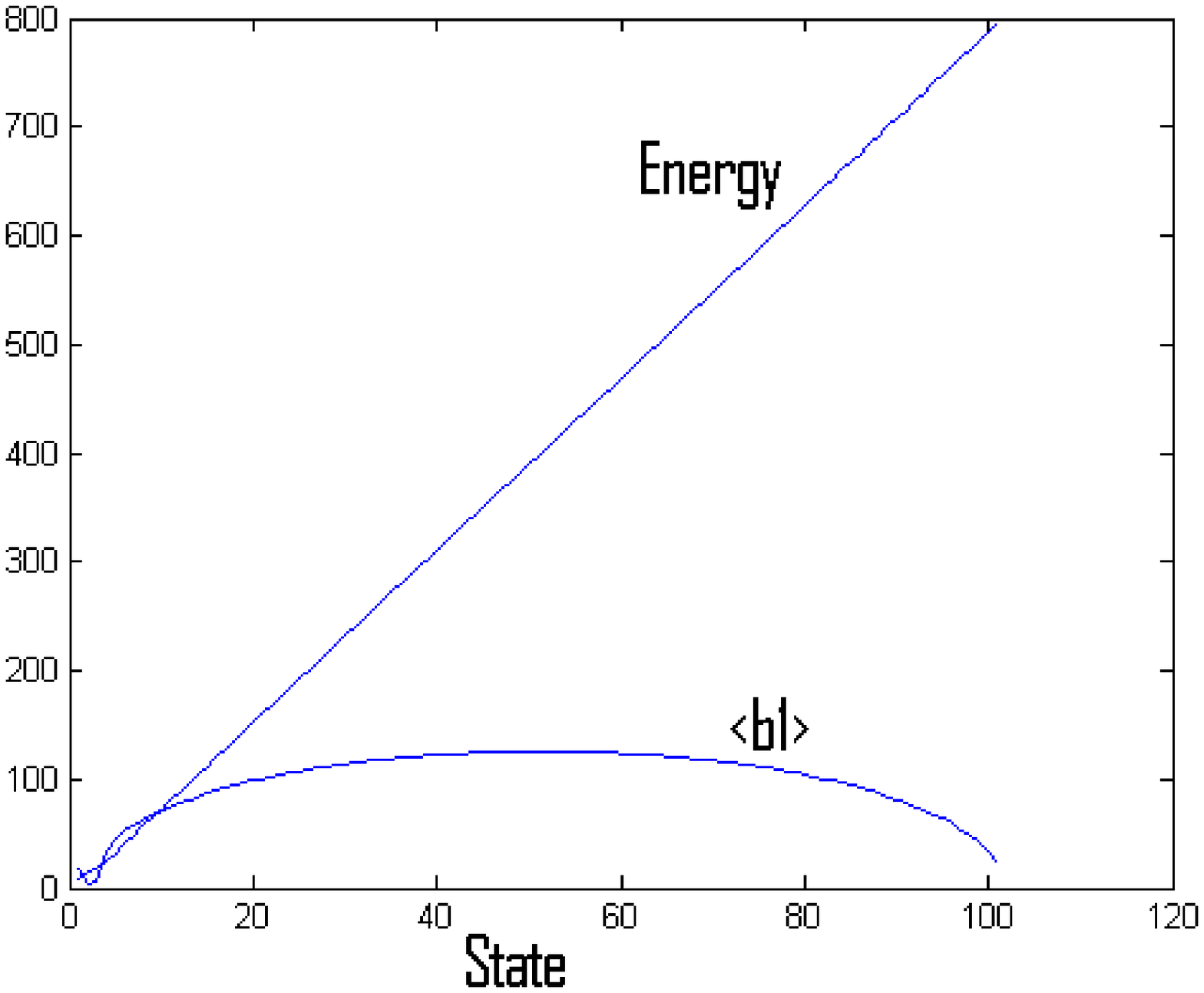}{10.0}{The spectrum of non-zero anomalous dimensions together with the expected value of $b_1$.
To produce this plot we used $N=100000$ and considered an operator built from $199800$ $Z$s and 2 $Y$s. There
are a total of 398 excited two threebrane states.}

The distance between the two threebranes is determined by the value of $b_1$. In figure 2 we have plotted the non-zero
energy eigenvalues together with the corresponding value of $\langle b_1\rangle$. Notice that for a given value of
$\langle b_1\rangle$ there are two energy eigenvalues. To interpret this note that each threebrane will
behave like a harmonic oscillator; the open strings stretched between these threebranes implies
that the oscillators are coupled. It is well known that coupled oscillators have two possible
normal modes, corresponding to the oscillators oscillating in phase or out of phase. The lowest energy state 
corresponds to the oscillators oscillating in phase; in this mode the strings stretching between the threebranes will
hardly be excited. When the oscillators oscillate out of phase, the strings stretching between the threebranes will
be excited making this the mode with the largest energy. When the threebranes are very close to each other, they
will join with ``dimples'' - they are too close for an actual open string to form. In this case, there is no open
string to excite which corresponds to the tiny ``V'' on the lower left hand side of the plot. When an open string 
has formed, the energy difference between the two states will be determined by how strongly the open string and 
the threebrane couple. This coupling \cite{Berenstein:2006qk,de Mello Koch:2007uv,Bekker:2007ea} has the form
$$\sqrt{1-{J\over N}}$$
where $J$ is the angular momentum of the giant. As the larger giant becomes nearly maximal, this coupling switches
off and the energy difference between the two states at fixed $\langle b_1\rangle$ decreases. The maximal giant is
decoupled (at one loop and at large $N$) from the open string. This is why, in figure 2 as $\langle b_1\rangle$ 
increases the difference in the energy of the two states decreases.

\section{Conclusions}

In this article we have determined the anomalous dimension of a class of operators
who classical dimension is $O(N)$. These operators have a dual interpretation as an excited two (threebrane)
state. An analogous computation for BMN loops allows one to see the worldsheet emerge as a lattice defined by
the matrices appearing in the loop. Our motivation was to see if an analogous result is possible for threebranes.
Our answer is clearly ``yes'' with the lattice emerging from the Young diagram labeling the restricted Schur 
polynomial. This lattice is a discretization of the radial coordinate of the threebrane - different lattice
points correspond to threebranes with different radii. Each threebrane behaves like a harmonic oscillator. The 
open strings stretched between these threebranes implies that the oscillators are coupled. Coupled oscillators 
have two possible normal modes, corresponding to the oscillators oscillating in phase or out of phase. These
two modes are evident from the numerically computed spectrum of anomalous dimensions. Further, the known 
strength of the coupling between the open string and the threebrane is clearly evident in the numerical spectrum.

By taking the angular momenta (that is $b_0$) of the threebranes to be at least $O(N)$, we have seen that in the large $N$
limit there is a dynamical decoupling so that the restricted Schur's with two columns do not mix with restricted Schur's
with a different number of columns. The number of columns can be identified with the number of threebranes, so that
the dual statement is simply that at weak string coupling the threebrane number is conserved. The fact that this
decoupling is achieved for certain values of the ${\cal R}$-charge is very similar to the BMN loop dynamics. The
${\cal R}$-charge of BMN loops $J$ is chosen to satisfy $J^2/N\ll 1$ to suppress mixing between single and multi traces.
We have also seen that the usual 't Hooft limit gives a well defined and non-trivial problem for the spectrum of the dilatation
operator.

There are a number of natural extensions of the present study. We would like to generalize the computation here to an
arbitrary number of $Y$ fields, and then even further by using many types of impurities. The computation of the present
paper has allowed a description of the radius of a giant graviton. It is only by considering the other complex Higgs scalars 
that we expect to see the worldvolume dimensions emerge. To consider general fluctuations of the giants it will be necessary
to include both the fermions and the gauge fields. Another natural question is if the 
dynamics for these excited threebranes is integrable or not.

{\vskip 1.0cm}

\noindent
{\it Acknowledgements:} We would like to thank Tom Brown, Tanay Dey, Dimitrios Giataganas, 
Yusuke Kimura, Sanjaye Ramgoolam, Shahin Sheikh-Jabbari and Dave Turton, for enjoyable, 
helpful discussions and/or correspondence. 
This work is based upon research supported by the South African Research Chairs
Initiative of the Department of Science and Technology and National Research Foundation.
Any opinion, findings and conclusions or recommendations expressed in this material
are those of the authors and therefore the NRF and DST do not accept any liability
with regard thereto.

\appendix

\section{Changing Basis, Character Identities}

\subsection{Changing the basis}

In these formulas $\{\cdot\}$ stand for labels that specify the Young diagram: the number
of boxes in the right most column (called $b_0$) and the number of boxes in
the left most column minus the number of boxes in the right most column (called $b_1$).
The weights\footnote{Recall that the weight of a box in row $i$ and
column $j$ is $N-i+j$.}
 $c_1,c_2$ are the weights of boxes $a$ and $b$ identified uniquely by the
requirement $c_1>c_2$. This fixes the weights to be
$$ 
c_1=N-b_0+1,\qquad c_2=N-b_0-b_1\, ,
$$
when {\bf the boxes are not in the same column}, or
$$ 
c_1=N-b_0+1,\qquad c_2=N-b_0\, ,
$$
when {\bf the boxes are both in the right most column}, or
$$ 
c_1=N-b_0-b_1,\qquad c_2=N-b_0-b_1-1\, , 
$$
when {\bf the boxes are both in the left most column}.

$$
|a;\{\cdot\}\rangle=
|{\tiny \yng(2,2,2,1,1,1)};{\tiny \yng(2,2,2,1)}\, i, {\tiny\yng(1,1)}\rangle =
|{\tiny\young({\,}{\,},{\,}{\,},{\,}{\,},{\,},{b},{a})}\, i\rangle 
=|1;\{\cdot\}\rangle,
$$
$$
|b;\{\cdot\}\rangle =
|{\tiny \yng(2,2,2,1,1)};{\tiny \yng(2,1,1,1,1)}\, i, {\tiny\yng(1,1)}\rangle =
|{\tiny\young({\,}{\,},{\,}{b},{\,}{a},{\,},{\,})}\, i\rangle 
=|2;\{\cdot\}\rangle,
$$
$$
|c;\{\cdot\}\rangle =
|{\tiny \yng(2,2,2,2,2)};{\tiny \yng(2,2,2,2)}\, i, {\tiny\yng(2)}\rangle =
|{\tiny\young({\,}{\,},{\,}{\,},{\,}{\,},{\,}{\,},{b}{a})}\, i\rangle =
|3;\{\cdot\}\rangle ,
$$
$$
|d;\{\cdot\}\rangle =
|{\tiny \yng(2,2,2,1,1)};{\tiny \yng(2,2,1,1)}\, i, {\tiny\yng(2)}\rangle =
\sqrt{c_1-c_2+1\over 2(c_1-c_2)}
|{\tiny\young({\,}{\,},{\,}{\,},{\,}{a},{\,},{b})}\, i\rangle
+ \sqrt{c_1-c_2-1 \over 2(c_1-c_2)} 
|{\tiny\young({\,}{\,},{\,}{\,},{\,}{b},{\,},{a})}\, i\rangle
$$
$$ =
\sqrt{c_1-c_2+1\over 2(c_1-c_2)}
|4;\{\cdot\}\rangle
+ \sqrt{c_1-c_2-1 \over 2(c_1-c_2)} |5;\{\cdot\}\rangle ,
$$
$$
|e;\{\cdot\}\rangle =
|{\tiny \yng(2,2,2,1,1)};{\tiny \yng(2,2,1,1)}\, i, {\tiny\yng(1,1)}\rangle =
\sqrt{c_1-c_2-1 \over 2(c_1-c_2)}
|{\tiny\young({\,}{\,},{\,}{\,},{\,}{a},{\,},{b})}\, i\rangle
-\sqrt{c_1-c_2+1\over 2(c_1-c_2)}
|{\tiny\young({\,}{\,},{\,}{\,},{\,}{b},{\,},{a})}\, i\rangle 
$$
$$
=\sqrt{c_1-c_2-1 \over 2(c_1-c_2)} |4;\{\cdot\}\rangle
-\sqrt{c_1-c_2+1\over 2(c_1-c_2)} |5;\{\cdot\}\rangle .
$$

\subsection{Character Identities}

The formulas of the previous subsection imply the following relations between characters
\begin{equation}
\chi_{{\tiny \yng(2,2,2,1,1,1)};{\tiny \yng(2,2,2,1)}\, {\tiny\yng(1,1)}}(\sigma) =
\chi_{\tiny\young({\,}{\,},{\,}{\,},{\,}{\,},{\,},{b},{a})}(\sigma ) 
\label{char1}
\end{equation}
\begin{equation}
\chi_{{\tiny \yng(2,2,2,1,1)};{\tiny \yng(2,1,1,1,1)}\, {\tiny\yng(1,1)}}(\sigma)=
\chi_{{\tiny\young({\,}{\,},{\,}{b},{\,}{a},{\,},{\,})}}(\sigma) 
\label{char2}
\end{equation}
\begin{equation}
\chi_{{\tiny \yng(2,2,2,2,2)};{\tiny \yng(2,2,2,2)}\, {\tiny\yng(2)}}(\sigma ) =
\chi_{{\tiny\young({\,}{\,},{\,}{\,},{\,}{\,},{\,}{\,},{b}{a})}}(\sigma )
\label{char3}
\end{equation}
\begin{equation}
\chi_{{\tiny \yng(2,2,2,1,1)};{\tiny \yng(2,2,1,1)}\, {\tiny\yng(2)}}(\sigma ) =
{b_1+2\over 2(b_1+1)}
\chi_{{\tiny\young({\,}{\,},{\,}{\,},{\,}{1},{\,},{2})}}(\sigma)
+ {b_1 \over 2(b_1+1)} 
\chi_{{\tiny\young({\,}{\,},{\,}{\,},{\,}{2},{\,},{1})}}(\sigma)
\label{char4}
\end{equation}
$$
+{\sqrt{b_1(b_1+2)}\over 2(b_1+1)}
\chi_{{\tiny\young({\,}{\,},{\,}{\,},{\,}{\onetwo},{\,},{\twoone})}}(\sigma)
+{\sqrt{b_1(b_1+2)}\over 2(b_1+1)}
\chi_{{\tiny\young({\,}{\,},{\,}{\,},{\,}{\twoone},{\,},{\onetwo})}}(\sigma)
$$
\begin{equation}
\chi_{{\tiny \yng(2,2,2,1,1)};{\tiny \yng(2,2,1,1)}\, {\tiny\yng(1,1)}}(\sigma ) =
{b_1\over 2(b_1+1)}
\chi_{{\tiny\young({\,}{\,},{\,}{\,},{\,}{1},{\,},{2})}}(\sigma)
+ {b_1 + 2\over 2(b_1+1)} 
\chi_{{\tiny\young({\,}{\,},{\,}{\,},{\,}{2},{\,},{1})}}(\sigma)
\label{char5}
\end{equation}
$$
-{\sqrt{b_1(b_1+2)}\over 2(b_1+1)}
\chi_{{\tiny\young({\,}{\,},{\,}{\,},{\,}{\onetwo},{\,},{\twoone})}}(\sigma)
-{\sqrt{b_1(b_1+2)}\over 2(b_1+1)}
\chi_{{\tiny\young({\,}{\,},{\,}{\,},{\,}{\twoone},{\,},{\onetwo})}}(\sigma)
$$

Finally, we will make use of the identities
$$
\chi_{\tiny \young({\,}{\,},{\,}{\,},{\,}{\,},{\,},{2},{1})}((n,n+2)\sigma -\sigma (n,n+2))
={1 \over b_1+3}\sqrt{1-{1\over (b_1+2)^2}}\left[
 \chi_{\tiny \young({\,}{\,},{\,}{\,},{\,}{\twothree},{\,},{\threetwo},{1})}(\sigma )
-\chi_{\tiny \young({\,}{\,},{\,}{\,},{\,}{\threetwo},{\,},{\twothree},{1})}(\sigma )
\right]
$$
$$
-\sqrt{1-{1\over (b_1+2)^2}}\sqrt{1-{1\over (b_1+3)^2}}\left[
 \chi_{\tiny \young({\,}{\,},{\,}{\,},{\,}{\onethree},{\,},{\threetwo},{\twoone})}(\sigma )
-\chi_{\tiny \young({\,}{\,},{\,}{\,},{\,}{\threeone},{\,},{\twothree},{\onetwo})}(\sigma )
\right]
$$
$$
\chi_{\tiny \young({\,}{\,},{\,}{2},{\,}{1},{\,},{\,},{\,})}((n,n+2)\sigma -\sigma (n,n+2))
=-{1 \over b_1-1}\sqrt{1-{1\over (b_1)^2}}\left[
 \chi_{\tiny \young({\,}{\,},{\,}{\threetwo},{\,}{1},{\,},{\,},{\twothree})}(\sigma )
-\chi_{\tiny \young({\,}{\,},{\,}{\twothree},{\,}{1},{\,},{\,},{\threetwo})}(\sigma )
\right]
$$
$$
-\sqrt{1-{1\over (b_1-1)^2}}\sqrt{1-{1\over (b_1)^2}}\left[
 \chi_{\tiny \young({\,}{\,},{\,}{\threetwo},{\,}{\twoone},{\,},{\,},{\onethree})}(\sigma )
-\chi_{\tiny \young({\,}{\,},{\,}{\twothree},{\,}{\onetwo},{\,},{\,},{\threeone})}(\sigma )
\right]
$$
$$
\chi_{\tiny \young({\,}{\,},{\,}{\,},{\,}{\,},{2}{1})}((n,n+2)\sigma -\sigma (n,n+2))
=-{\sqrt{3}\over 2}\left[
 \chi_{\tiny \young({\,}{\,},{\,}{\,},{\,}{\twothree},{\threetwo}{1})}(\sigma)
-\chi_{\tiny \young({\,}{\,},{\,}{\,},{\,}{\threetwo},{\twothree}{1})}(\sigma)
\right]
$$
$$
\chi_{\tiny \young({\,}{\,},{\,}{\,},{\,}{2},{\,},{\,},{1})}((n,n+2)\sigma -\sigma (n,n+2))
={1 \over b_1+1}\sqrt{1-{1\over (b_1)^2}}\left[
 \chi_{\tiny \young({\,}{\,},{\,}{\,},{\,}{\threetwo},{\,},{\twothree},{1})}(\sigma )
-\chi_{\tiny \young({\,}{\,},{\,}{\,},{\,}{\twothree},{\,},{\threetwo},{1})}(\sigma )
\right]
$$
$$
-{1 \over b_1}\sqrt{1-{1\over (b_1+1)^2}}\left[
 \chi_{\tiny \young({\,}{\,},{\,}{\,},{\,}{\onetwo},{\,},{3},{\twoone})}(\sigma )
-\chi_{\tiny \young({\,}{\,},{\,}{\,},{\,}{\twoone},{\,},{3},{\onetwo})}(\sigma )
\right]
$$
$$
+{1\over b_1+2}\sqrt{1-{1\over (b_1+1)^2}}\left[
 \chi_{\tiny \young({\,}{\,},{\,}{3},{\,}{\onetwo},{\,},{\,},{\twoone})}(\sigma )
-\chi_{\tiny \young({\,}{\,},{\,}{3},{\,}{\twoone},{\,},{\,},{\onetwo})}(\sigma )
\right]
$$
$$
-\sqrt{1-{1\over (b_1+1)^2}}\sqrt{1-{1\over (b_1+2)^2}}\left[
 \chi_{\tiny \young({\,}{\,},{\,}{\twothree},{\,}{\onetwo},{\,},{\,},{\threeone})}(\sigma )
-\chi_{\tiny \young({\,}{\,},{\,}{\threetwo},{\,}{\twoone},{\,},{\,},{\onethree})}(\sigma )
\right]
$$
$$
\chi_{\tiny \young({\,}{\,},{\,}{\,},{\,}{1},{\,},{\,},{2})}((n,n+2)\sigma -\sigma (n,n+2))
=-{1 \over b_1}\sqrt{1-{1\over (b_1+1)^2}}\left[
 \chi_{\tiny \young({\,}{\,},{\,}{\,},{\,}{\twoone},{\,},{3},{\onetwo})}(\sigma )
-\chi_{\tiny \young({\,}{\,},{\,}{\,},{\,}{\onetwo},{\,},{3},{\twoone})}(\sigma )
\right]
$$
$$
-\sqrt{1-{1 \over (b_1)^2}}\sqrt{1-{1\over (b_1+1)^2}}\left[
 \chi_{\tiny \young({\,}{\,},{\,}{\,},{\,}{\threeone},{\,},{\twothree},{\onetwo})}(\sigma )
-\chi_{\tiny \young({\,}{\,},{\,}{\,},{\,}{\onethree},{\,},{\threetwo},{\twoone})}(\sigma )
\right]
$$
$$
+{1\over b_1+1}\sqrt{1-{1\over (b_1+2)^2}}\left[
 \chi_{\tiny \young({\,}{\,},{\,}{\twothree},{\,}{1},{\,},{\,},{\threetwo})}(\sigma )
-\chi_{\tiny \young({\,}{\,},{\,}{\threetwo},{\,}{1},{\,},{\,},{\twothree})}(\sigma )
\right]
$$
$$
+{1\over b_1+2}\sqrt{1-{1\over (b_1+1)^2}}\left[
 \chi_{\tiny \young({\,}{\,},{\,}{3},{\,}{\onetwo},{\,},{\,},{\threeone})}(\sigma )
-\chi_{\tiny \young({\,}{\,},{\,}{3},{\,}{\twoone},{\,},{\,},{\onethree})}(\sigma )
\right]
$$

\subsection{New Identities}

In this subsection we will derive identities that will allow us to
express terms involving a restricted Schur polynomial in $Y^2$ or a 
trace of $Y$ times a restricted Schur polynomial, as a linear combination 
of restricted Schur polynomials in $Y$. Start from the expression
$$\chi_{R',R''} (Z,Y^2)={1\over n!}\sum_{\sigma\in S_{n+1}}\chi_{R',R''}(\sigma )Z^{i_1}_{i_{\sigma (1)}}\cdots
Z^{i_n}_{i_{\sigma (n)}}(Y^2)^{i_{n+1}}_{i_{\sigma (n+1)}}$$
$$
={1\over n!}\sum_{\sigma\in S_{n+1}}\chi_{R',R''}(\sigma )Z^{i_1}_{i_{\sigma (1)}}\cdots
Z^{i_n}_{i_{\sigma (n)}}Y^{i_{n+1}}_{i_{n+2}}Y^{i_{n+2}}_{i_{\sigma (n+1)}}
$$
$$
={1\over n!}\sum_{\sigma\in S_{n+1}}\chi_{R',R''}(\sigma ){\rm Tr}(\sigma (n+1,n+2)Z^{\otimes\, n} Y^{\otimes\, 2})
$$
$$
={1\over n!}\sum_{\sigma\in S_{n+1}}\chi_{R',R''}(\sigma )\sum_{R,(r_1,r_2)}\chi^{R,(r_1,r_2)}(\sigma (n+1,n+2))
\chi_{R,(r_1,r_2)}(Z , Y)
$$
$$
=\sum_{R,(r_1,r_2)}\alpha_{R,(r_1,r_2)}\chi_{R,(r_1,r_2)}(Z , Y)\, .
$$
In the above, $\chi^{R,(r_1,r_2)}(\sigma (n+1,n+2))$ is the dual character, computed and defined in \cite{Bhattacharyya:2008rc}. It is given by
$$
\chi^{R,(r_1,r_2)}(\sigma (n+1,n+2))={d_R n! 2!\over d_{r_1}d_{r_2}(n+2)!}\chi_{R,(r_1,r_2)}(\sigma (n+1,n+2))\, .
$$
The coefficients $\alpha_{R,(r_1,r_2)}$ are
$$
\alpha_{R,(r_1,r_2)} ={1\over n!}
\sum_{\sigma\in S_{n+1}}\chi_{R',R''}(\sigma )\chi^{R,(r_1,r_2)}(\sigma (n+1,n+2))
$$
$$
={d_R 2!\over d_{r_1}d_{r_2}(n+2)!}
\sum_{\sigma\in S_{n+1}}\chi_{R',R''}(\sigma )\chi_{R,(r_1,r_2)}(\sigma (n+1,n+2))\, .
$$
Lets do the sum over $\sigma$ by rewriting this last expression as
$$
\alpha_{R,(r_1,r_2)} ={d_R 2!\over d_{r_1}d_{r_2}(n+2)!}
{\rm Tr}_{R,(r_1,r_2)}\left( \sum_{\sigma\in S_{n+1}}\chi_{R',R''}(\sigma ) \sigma (n+1,n+2)\right)\, .
$$
The reason why this is a useful rewriting is that it is easy to recognize that
$$
\sum_{\sigma\in S_{n+1}}\chi_{R',R''}(\sigma ) \sigma = {(n+1)!\over d_{R'}}P_{R',R''}
$$
is $P_{R',R''}$, a projection operator acting on $R'$, projecting to
$R''$. To see that the LHS is a projector, the reader is encouraged to
verify that, for example,  it squares to itself and that if one sums over $R'$ and 
$R''$, one recovers the standard projector onto irreducible representation $R$. Finally, 
she can also check that it annihilates states not in the $R''$ subspace. Thus, we obtain (use $d_{r_2}=1$)
$$
\alpha_{R,(r_1,r_2)} = {2 {\rm hooks}_{R'}\over {\rm hooks}_{R}d_{r_1}}\sum_{i,j}
\langle R,(r_1,r_2); i |R',R''; j\rangle\langle R',R''; j|(n+1,n+2)|R,(r_1,r_2);i\rangle
$$
which is a lovely explicit expression that is easy to evaluate.

It is equally easy to argue that
$$
\chi_{R',R''} (Z,Y){\rm Tr}(Y)=\sum_{R,(r_1,r_2)}\beta_{R,(r_1,r_2)}\chi_{R,(r_1,r_2)}(Z , Y)\, ,
$$
where
$$
\beta_{R,(r_1,r_2)} = {2 {\rm hooks}_{R'}\over {\rm hooks}_{R}d_{r_1}}\sum_{i,j}
\langle R,(r_1,r_2); i |R',R''; j\rangle\langle R',R''; j|R,(r_1,r_2);i\rangle\, .
$$
Applying these formulas we obtain (the box with a star on the LHS contains the $Y^2$)
$$
\chi_{\tiny \young({\,}{\,},{\,}{\,},{\,}{*},{\,},{\,},{\,})}= 
{1\over b_0+b_1+2}\chi_{\tiny \young({\,}{\,},{\,}{\,},{\,}{\,},{\,},{\,},{\,},{\,}); \young({\,}{\,},{\,}{\,},{\,},{\,},{\,},{\,})\yng(2)}
-{b_1+2\over b_1(b_0+b_1+2)}\chi_{\tiny \young({\,}{\,},{\,}{\,},{\,}{\,},{\,},{\,},{\,},{\,}); \young({\,}{\,},{\,}{\,},{\,},{\,},{\,},{\,}),\yng(1,1)}
-{2(b_1-1)\over b_1(b_0+2)}\chi_{\tiny \young({\,}{\,},{\,}{\,},{\,}{\,},{\,}{\,},{\,},{\,}); \young({\,}{\,},{\,}{\,},{\,},{\,},{\,},{\,}),\yng(1,1)}
$$
$$ 
+{b_0+b_1+1\over b_0+b_1+2}\chi_{\tiny \young({\,}{\,}{\,},{\,}{\,},{\,}{\,},{\,},{\,},{\,}); \young({\,}{\,},{\,}{\,},{\,},{\,},{\,},{\,})\yng(2)}
-{b_0(b_0+b_1+1)\over (b_0+2)(b_0+b_1+2)}\chi_{\tiny \young({\,}{\,}{\,},{\,}{\,},{\,}{\,},{\,},{\,},{\,}); \young({\,}{\,},{\,}{\,},{\,},{\,},{\,},{\,})\yng(1,1)}
$$
and (the box with the star on the LHS contains the $Y$)
$$
{\rm Tr}(Y)\chi_{\tiny \young({\,}{\,},{\,}{\,},{\,}{*},{\,},{\,},{\,})}= 
{1\over b_0+b_1+2}\chi_{\tiny \young({\,}{\,},{\,}{\,},{\,}{\,},{\,},{\,},{\,},{\,}); \young({\,}{\,},{\,}{\,},{\,},{\,},{\,},{\,})\yng(2)}
+{b_1+2\over b_1(b_0+b_1+2)}\chi_{\tiny \young({\,}{\,},{\,}{\,},{\,}{\,},{\,},{\,},{\,},{\,}); \young({\,}{\,},{\,}{\,},{\,},{\,},{\,},{\,}),\yng(1,1)}
+{2(b_1-1)\over b_1(b_0+2)}\chi_{\tiny \young({\,}{\,},{\,}{\,},{\,}{\,},{\,}{\,},{\,},{\,}); \young({\,}{\,},{\,}{\,},{\,},{\,},{\,},{\,}),\yng(1,1)}
$$
$$ 
+{b_0+b_1+1\over b_0+b_1+2}\chi_{\tiny \young({\,}{\,}{\,},{\,}{\,},{\,}{\,},{\,},{\,},{\,}); \young({\,}{\,},{\,}{\,},{\,},{\,},{\,},{\,})\yng(2)}
+{b_0(b_0+b_1+1)\over (b_0+2)(b_0+b_1+2)}\chi_{\tiny \young({\,}{\,}{\,},{\,}{\,},{\,}{\,},{\,},{\,},{\,}); \young({\,}{\,},{\,}{\,},{\,},{\,},{\,},{\,})\yng(1,1)}
$$
and (the box with the star on the LHS contains the $Y^2$)
$$\chi_{\tiny\young({\,}{\,},{\,}{\,},{\,}{\,},{\,},{\,},{\,},{*})}=
{1\over b_0+1}\chi_{\tiny \yng(2,2,2,2,1,1,1);\yng(2,2,2,1,1,1)\yng(2)}
-{b_1\over (b_0+1)(b_1+2)}\chi_{\tiny \yng(2,2,2,2,1,1,1);\yng(2,2,2,1,1,1)\yng(1,1)}
-{2(b_1+3)\over (b_1+2)(b_0+b_1+3)}
\chi_{\tiny \yng(2,2,2,1,1,1,1,1)\yng(2,2,2,1,1,1)\yng(1,1)}
$$
$$
+{b_0\over b_0+1}\chi_{\tiny\yng(3,2,2,1,1,1,1);\yng(2,2,2,1,1,1)\yng(2)}
-{b_0\over b_0+1}{b_0+b_1+1\over b_0+b_1+3}
\chi_{\tiny\yng(3,2,2,1,1,1,1);\yng(2,2,2,1,1,1)\yng(1,1)}
$$
and (the box with the star on the LHS contains the $Y$)
$${\rm Tr}(Y)\chi_{\tiny\young({\,}{\,},{\,}{\,},{\,}{\,},{\,},{\,},{\,},{*})}=
{1\over b_0+1}\chi_{\tiny \yng(2,2,2,2,1,1,1);\yng(2,2,2,1,1,1)\yng(2)}
+{b_1\over (b_0+1)(b_1+2)}\chi_{\tiny \yng(2,2,2,2,1,1,1);\yng(2,2,2,1,1,1)\yng(1,1)}
+{2(b_1+3)\over (b_1+2)(b_0+b_1+3)}
\chi_{\tiny \yng(2,2,2,1,1,1,1,1)\yng(2,2,2,1,1,1)\yng(1,1)}
$$
$$
+{b_0\over b_0+1}\chi_{\tiny\yng(3,2,2,1,1,1,1);\yng(2,2,2,1,1,1)\yng(2)}
+{b_0\over b_0+1}{b_0+b_1+1\over b_0+b_1+3}
\chi_{\tiny\yng(3,2,2,1,1,1,1);\yng(2,2,2,1,1,1)\yng(1,1)}\, .
$$
These are all of the identities that we will need.

\section{Normalization Factors}

Normalized operators are normalized to have a unit two point function. Thus, to compute the normalization factor of any operator we simply 
need to compute its two point function. In this appendix we will compute the two point functions of all operators that appear.
We use the notation $f_R$ to denote the product of all the weights in Young diagram $R$. Further, we list the row lengths to specify
the Young diagram. Thus, for example, $f_{(3,2^{b_0-1},1^{b_1+1})}$ is the product of weights in the Young diagram that has one
row with 3 boxes in it, $b_0-1$ rows with 2 boxes and $b_1+1$ rows with a single box. The two point functions we need are

$$
\left\langle O_a(b_0,b_1)O_a^\dagger (b_0,b_1)\right\rangle = f_{(2^{b_0},1^{b_1+2})}{(b_1+1)(b_0+b_1+2)(b_0+b_1+3)\over 2(b_1+3)}
$$

$$
\left\langle O_b(b_0,b_1)O_b^\dagger (b_0,b_1)\right\rangle = f_{(2^{b_0+2},1^{b_1-2})}{(b_1+1)(b_0+1)(b_0+2)\over 2(b_1-1)}
$$

$$
\left\langle O_c(b_0,b_1)O_c^\dagger (b_0,b_1)\right\rangle = f_{(2^{b_0+1})}{(b_0+1)(b_0+2)\over 2}
$$

$$
\left\langle O_d(b_0,b_1)O_d^\dagger (b_0,b_1)\right\rangle = f_{(2^{b_0+1},1^{b_1})}{(b_0+1)(b_0+b_1+2)\over 2}=\left\langle O_e(b_0,b_1)O_e^\dagger (b_0,b_1)\right\rangle
$$

$$
\left\langle O_f(b_0,b_1)O_f^\dagger (b_0,b_1)\right\rangle = f_{(3,2^{b_0-1},1^{b_1+1})}{(b_0+1)(b_0+b_1+3)(b_1+1)\over 2b_0(b_1+2)}
=\left\langle O_g(b_0,b_1)O_g^\dagger (b_0,b_1)\right\rangle
$$

$$
\left\langle O_h(b_0,b_1)O_h^\dagger (b_0,b_1)\right\rangle = f_{(3,2^{b_0},1^{b_1-1})}{(b_0+2)(b_0+b_1+2)(b_1+1)\over 2b_1(b_0+b_1+1)}
=\left\langle O_i(b_0,b_1)O_i^\dagger (b_0,b_1)\right\rangle
$$

In the limit we consider $O_f$, $O_g$, $O_h$ and $O_i$ are all subleading. This is crucial to show that we do indeed have 
a subsector that decouples dynamically.

\section{Exact Action of the One Loop Dilatation Operator}

The action of the dilatation operator given in this section was checked by evaluating both sides numerically
with randomly generated $Z$ and $Y$, for the case that $R$ contains 5 or 6 boxes.

$$ 
DO_a (b_0,b_1)=
4{(N-b_0-b_1-1)\over (b_1+2)^2}\left(1
-{b_1+3\over b_0+b_1+3}
\right)O_a(b_0,b_1)
$$
$$
-2{(N-b_0-b_1-1)\over b_1+2}\left(1
-{1\over b_0+1}\right)
O_d(b_0,b_1)
$$
$$
+2{(N-b_0-b_1-1)b_1\over (b_1+2)^2}\left( 1 -{1\over b_0+1}
\right) O_e(b_0,b_1)
$$
$$
+4{(b_1+1)(N-b_0-b_1-1)\over (b_1+3)(b_1+2)^2}\left(1
+{(b_1+1)\over (b_0+1)} \right) O_b(b_0-1,b_1+2)
$$
$$
+2{(b_1+1)(N-b_0-b_1-1)\over (b_1+2)(b_1+3)}\left(1
-{1\over b_0+b_1+3}
\right)O_d(b_0-1,b_1+2)
$$
$$
-2{(b_1+4)(b_1+1)(N-b_0-b_1-1)\over (b_1+3)(b_1+2)^2}\left(
1-{1\over (b_0+b_1+3)}\right)O_e(b_0-1,b_1+2)
$$
$$
+2{(N-b_0-b_1-1)\over b_1+2}{b_0\over b_0+1}O_f (b_0,b_1)
-2{(N-b_0-b_1-1)\over b_1+2}{b_0\over b_0+1}{b_0+b_1+1\over b_0+b_1+3}
O_g(b_0,b_1)
$$
$$ 
-2{(N-b_0-b_1-1)(b_1+1)\over (b_1+2)(b_1+3)}
{b_0+b_1+2\over b_0+b_1+3}O_h(b_0-1,b_1+2)
$$
$$
+2{(N-b_0-b_1-1)(b_1+1)\over (b_1+2)(b_1+3)}{(b_0-1)(b_0+b_1+2)\over (b_0+1)(b_0+b_1+3)}O_i(b_0-1,b_1+2)\, .
$$

$$ $$

$$
D O_b(b_0,b_1) =
4{(N-b_0)(b_1+1)\over (b_1-1)b_1^2}\left(1 - {b_1+1\over b_0+b_1+2}\right) O_a(b_0+1,b_1-2)
$$
$$
-2{(N-b_0)(b_1+1)\over b_1 (b_1-1)}\left(1-{1\over b_0+2}\right)O_d(b_0+1,b_1-2)
$$
$$
+2{(N-b_0)(b_1+1)(b_1-2)\over b_1^2 (b_1-1)}\left(1-{1\over b_0+2}\right) O_e(b_0+1,b_1-2)
$$
$$
+4{(N-b_0)\over b_1^2}\left(1+{b_1-1\over b_0+2}\right)O_b(b_0,b_1)
$$
$$
+2{N-b_0\over b_1}\left(1-{1\over b_0+b_1+2}\right)O_d(b_0,b_1)
$$
$$
-2{(b_1+2)(N-b_0)\over b_1^2}\left( 1-{1\over b_0+b_1+2}\right) O_e(b_0,b_1)
$$
$$
+2{(N-b_0)(b_1+1)\over b_1(b_1-1)}{b_0+1\over b_0+2}O_f(b_0+1,b_1-2)
-2{(N-b_0)(b_1+1)\over b_1(b_1-1)}{b_0+1\over b_0+2}{b_0+b_1\over b_0+b_1+2}O_g(b_0+1,b_1-2)
$$
$$
-2{N-b_0\over b_1}{b_0+b_1+1\over b_0+b_1+2}O_h(b_0,b_1) 
+2{N-b_0\over b_1}{b_0(b_0+b_1+1)\over (b_0+2)(b_0+b_1+2)}O_i(b_0,b_1)\, .
$$

$$ $$

$$
DO_d(b_0,b_1)=
-2{(b_1+3)(N-b_0+1)\over (b_1+1)(b_1+2)}\left(1-{b_1+3\over b_0+b_1+3}\right) O_a(b_0,b_1)
$$
$$
+2{(b_1-1)(N-b_0-b_1)\over b_1(b_1+1)}\left(
1+{b_1-1\over b_0+2}
\right) O_b(b_0,b_1)
$$
$$
+\left( 2N-2b_0-b_1+3 -{(N-b_0+1)(b_1+3)\over (b_0+1)(b_1+1)}-{(b_1-1)(N-b_0-b_1)\over (b_1+1)(b_0+b_1+2)}\right)O_d(b_0,b_1)
$$
$$
+\left({(N-b_0)(4-4b_1-2b_1^2)+b_1^3+b_1^2-4b_1\over b_1(b_1+2)}
+{b_1(b_1+3)(N-b_0+1)\over (b_0+1)(b_1+1)(b_1+2)}\right.$$
$$
\left.
+{(b_1-1)(N-b_0-b_1)(b_1+2)\over (b_1+1)b_1(b_0+b_1+2)}\right)O_e(b_0,b_1)
$$
$$
-2{(N-b_0+1)\over b_1+2}\left(1+{b_1+1\over b_0+1}\right)O_b(b_0-1,b_1+2)
$$
$$
-(N-b_0+1)\left(1-{1\over b_0+b_1+3}\right)O_d(b_0-1,b_1+2)
$$
$$
+{(N-b_0+1)(b_1+4)\over (b_1+2)}\left(1-{1\over b_0+b_1+3}\right)O_e(b_0-1,b_1+2)
$$
$$
+2{(N-b_0-b_1)\over b_1}\left(1-{b_1+1\over b_0+b_1+2}
\right)O_a(b_0+1,b_1-2)
$$
$$
-(N-b_0-b_1)\left(1-{1\over b_0+2}\right)O_d(b_0+1,b_1-2)
$$
$$
+{(b_1-2)(N-b_0-b_1)\over b_1}\left(1-{1\over b_0+2}\right)O_e(b_0+1,b_1-2)
$$
$$
-{(b_1+3)(N-b_0+1)\over b_1+1}{b_0\over b_0+1}O_f(b_0,b_1)
+{(b_1+3)(N-b_0+1)\over b_1+1}{b_0\over b_0+1}{b_0+b_1+1\over b_0+b_1+3}O_g(b_0,b_1)
$$
$$
+(N-b_0+1){b_0+b_1+2\over b_0+b_1+3}O_h(b_0-1,b_1+2)
-(N-b_0+1){(b_0-1)(b_0+b_1+2)\over (b_0+1)(b_0+b_1+3)}O_i(b_0-1,b_1+2)
$$
$$
-{(b_1-1)(N-b_0-b_1)\over (b_1+1)}{b_0+b_1+1\over b_0+b_1+2}O_h(b_0,b_1)
+{(b_1-1)(N-b_0-b_1)\over b_1+1}{b_0(b_0+b_1+1)\over (b_0+2)(b_0+b_1+2)}O_i(b_0,b_1)
$$
$$
+(N-b_0-b_1){b_0+1\over b_0+2}O_f(b_0+1,b_1-2)-
(N-b_0-b_1){b_0+1\over b_0+2}{b_0+b_1\over b_0+b_1+2}O_g(b_0+1,b_1-2)
$$

$$ $$

$$
D O_e(b_0,b_1)=
2{(b_1+3)b_1(N-b_0+1)\over (b_1+1)(b_1+2)^2}
\left(1-{b_1+3\over b_0+b_1+3} \right)O_a(b_0,b_1)
$$
$$
-2{(b_1-1)(b_1+2)(N-b_0-b_1)\over b_1^2(b_1+1)}\left(
1+{b_1-1\over b_0+2}
\right)O_b(b_0,b_1)
$$
$$
+\left({(N-b_0)(4 -2b_1^2 -4b_1)+b_1^3+b_1^2-4b_1\over b_1(b_1+2)}+{(b_1+3)(N-b_0+1)b_1\over (b_0+1)(b_1+1)(b_1+2)}\right.
$$
$$
\left. +{(b_1-1)(b_1+2)(N-b_0-b_1)\over b_1(b_1+1)(b_0+b_1+2)}\right)O_d(b_0,b_1)
$$
$$
+\left({2(N-b_0)(b_1^4 + 4b_1^3 + 4b_1^2 - 8)- b_1^5 -5 b_1^4 -8b_1^3 + 16b_1\over b_1^2(b_1+2)^2}
-{b_1^2(b_1+3)(N-b_0+1)\over (b_0+1)(b_1+1)(b_1+2)^2}\right.
$$
$$
\left.
-{(b_1-1)(b_1+2)(N-b_0-b_1)(b_1+2)\over b_1^2 (b_1+1)(b_0+b_1+2)}
\right)O_e(b_0,b_1)
$$
$$
+2{(N-b_0+1)b_1\over (b_1+2)^2}\left(
1+{b_1+1 \over b_0+1}
\right)O_b(b_0-1,b_1+2)
$$
$$
+{(N-b_0+1)b_1\over (b_1+2)}\left(
1-{1\over b_0+b_1+3}
\right) O_d(b_0-1,b_1+2)
$$
$$
-{(N-b_0+1)b_1(b_1+4)\over (b_1+2)^2}\left(
1-{1\over b_0+b_1+3}
\right) O_e(b_0-1,b_1+2)
$$
$$
-2{(N-b_0-b_1)(b_1+2)\over b_1^2}\left(
1-{b_1+1\over b_0+b_1+2}
\right) O_a(b_0+1,b_1-2)
$$
$$
+{(N-b_0-b_1)(b_1+2)\over b_1}\left(
1-{1\over b_0+2}
\right) O_d(b_0+1,b_1-2)
$$
$$
-{(b_1+2)(b_1-2)(N-b_0-b_1)\over b_1^2}\left(
1-{1\over b_0+2}
\right)O_e(b_0+1,b_1-2)
$$
$$
+{(b_1+3)(N-b_0+1)b_1\over (b_1+1)(b_1+2)}{b_0\over b_0+1}O_f(b_0,b_1)
-{(b_1+3)(N-b_0+1)b_1\over (b_1+1)(b_1+2)}{b_0(b_0+b_1+1)\over (b_0+1)(b_0+b_1+3)}O_g(b_0,b_1)
$$
$$
-{b_1 (N-b_0+1)\over b_1+2}{b_0+b_1+2 \over b_0+b_1+3}O_h(b_0-1,b_1+2)
$$
$$
+{b_1 (N-b_0+1)\over b_1+2}{(b_0+b_1+2)(b_0-1) \over (b_0+b_1+3)(b_0+1)}O_i(b_0-1,b_1+2)
$$
$$
+{(b_1-1)(b_1+2)(N-b_0-b_1)\over b_1(b_1+1)}{b_0+b_1+1\over b_0+b_1+2} O_h(b_0,b_1)
$$
$$
-{(b_1-1)(b_1+2)(N-b_0-b_1)b_0(b_0+b_1+1)\over b_1(b_1+1)(b_0+2)(b_0+b_1+2)}O_i(b_0,b_1)
$$
$$
-{(b_1+2)(N-b_0-b_1)\over b_1}{b_0+1\over b_0+2}O_f(b_0+1,b_1-2)
$$
$$
+{(b_1+2)(N-b_0-b_1)\over b_1}{b_0+1\over b_0+2}{b_0+b_1\over b_0+b_1+2}O_g(b_0+1,b_1-2)
$$

These formulas are exact. They will, of course, simplify dramatically once we drop subleading terms.


\begin{thebibliography}{30}
\parskip-2pt

\bibitem{Maldacena:1997re}
  J.~M.~Maldacena,
  ``The large N limit of superconformal field theories and supergravity,''
  Adv.\ Theor.\ Math.\ Phys.\  {\bf 2}, 231 (1998)
  [Int.\ J.\ Theor.\ Phys.\  {\bf 38}, 1113 (1999)]
  [arXiv:hep-th/9711200];\\
  S.~S.~Gubser, I.~R.~Klebanov and A.~M.~Polyakov,
  ``Gauge theory correlators from non-critical string theory,''
  Phys.\ Lett.\ B {\bf 428}, 105 (1998)
  [arXiv:hep-th/9802109];\\
  E.~Witten,
  ``Anti-de Sitter space and holography,''
  Adv.\ Theor.\ Math.\ Phys.\  {\bf 2}, 253 (1998)
  [arXiv:hep-th/9802150].

\bibitem{'tHooft:1973jz}
  G.~'t Hooft,
  ``A Planar Diagram Theory for Strong Interactions,''
  Nucl.\ Phys.\  B {\bf 72}, 461 (1974);\\
  G.~'t Hooft,
  ``A Two-Dimensional Model For Mesons,''
  Nucl.\ Phys.\  B {\bf 75}, 461 (1974).

\bibitem{Berenstein:2002jq}
  D.~E.~Berenstein, J.~M.~Maldacena and H.~S.~Nastase,
  ``Strings in flat space and pp waves from N = 4 super Yang Mills,''
  JHEP {\bf 0204}, 013 (2002)
  [arXiv:hep-th/0202021].

\bibitem{Kruczenski:2003gt}
  M.~Kruczenski,
  ``Spin chains and string theory,''
  Phys.\ Rev.\ Lett.\  {\bf 93}, 161602 (2004)
  [arXiv:hep-th/0311203],\\
  M.~Kruczenski, A.~V.~Ryzhov and A.~A.~Tseytlin,
  ``Large spin limit of AdS(5) x S**5 string theory and low energy  expansion
    of ferromagnetic spin chains,''
  Nucl.\ Phys.\  B {\bf 692}, 3 (2004)
  [arXiv:hep-th/0403120].

\bibitem{Hernandez:2004uw}
  R.~Hernandez and E.~Lopez,
  ``The SU(3) spin chain sigma model and string theory,''
  JHEP {\bf 0404}, 052 (2004)
  [arXiv:hep-th/0403139].

\bibitem{Bellucci:2004qr}
  S.~Bellucci, P.~Y.~Casteill, J.~F.~Morales and C.~Sochichiu,
  ``SL(2) spin chain and spinning strings on AdS(5) x S**5,''
  Nucl.\ Phys.\  B {\bf 707}, 303 (2005)
  [arXiv:hep-th/0409086].

\bibitem{Polchinski:1995mt}
  J.~Polchinski,
  ``Dirichlet-Branes and Ramond-Ramond Charges,''
  Phys.\ Rev.\ Lett.\  {\bf 75}, 4724 (1995)
  [arXiv:hep-th/9510017].

\bibitem{McGreevy:2000cw}
  J.~McGreevy, L.~Susskind and N.~Toumbas,
  ``Invasion of the giant gravitons from anti-de Sitter space,''
  JHEP {\bf 0006}, 008 (2000)
  [arXiv:hep-th/0003075];\\
  M.~T.~Grisaru, R.~C.~Myers and O.~Tafjord,
  ``SUSY and Goliath,''
  JHEP {\bf 0008}, 040 (2000)
  [arXiv:hep-th/0008015];\\
  A.~Hashimoto, S.~Hirano and N.~Itzhaki,
  ``Large branes in AdS and their field theory dual,''
  JHEP {\bf 0008}, 051 (2000)
  [arXiv:hep-th/0008016].

\bibitem{Balasubramanian:2001nh}
  V.~Balasubramanian, M.~Berkooz, A.~Naqvi and M.~J.~Strassler,
  ``Giant gravitons in conformal field theory,''
  JHEP {\bf 0204}, 034 (2002)
  [arXiv:hep-th/0107119].

\bibitem{Corley:2001zk}
  S.~Corley, A.~Jevicki and S.~Ramgoolam,
  ``Exact correlators of giant gravitons from dual N = 4 SYM theory,''
  Adv.\ Theor.\ Math.\ Phys.\  {\bf 5}, 809 (2002)
  [arXiv:hep-th/0111222].

\bibitem{Berenstein:2004kk}
  D.~Berenstein,
  ``A toy model for the AdS/CFT correspondence,''
  JHEP {\bf 0407}, 018 (2004)
  [arXiv:hep-th/0403110].

\bibitem{Balasubramanian:2004nb}
  V.~Balasubramanian, D.~Berenstein, B.~Feng and M.~x.~Huang,
  ``D-branes in Yang-Mills theory and emergent gauge symmetry,''
  JHEP {\bf 0503}, 006 (2005)
  [arXiv:hep-th/0411205].

\bibitem{de Mello Koch:2007uu}
  R.~de Mello Koch, J.~Smolic and M.~Smolic,
  ``Giant Gravitons - with Strings Attached (I),'' JHEP {\bf 0706}, 074 (2007),
  arXiv:hep-th/0701066.

\bibitem{de Mello Koch:2007uv}
  R.~de Mello Koch, J.~Smolic and M.~Smolic,
  ``Giant Gravitons - with Strings Attached (II),'' JHEP {\bf 0709} 049 (2007),
  arXiv:hep-th/0701067.

\bibitem{Bekker:2007ea}
  D.~Bekker, R.~de Mello Koch and M.~Stephanou,
  ``Giant Gravitons - with Strings Attached (III),''
  arXiv:0710.5372 [hep-th].

\bibitem{Bhattacharyya:2008rb}
  R.~Bhattacharyya, S.~Collins and R.~d.~M.~Koch,
  ``Exact Multi-Matrix Correlators,''
  JHEP {\bf 0803}, 044 (2008)
  [arXiv:0801.2061 [hep-th]].

\bibitem{Kimura:2008wy}
 Y.~Kimura and S.~Ramgoolam,
  ``Enhanced symmetries of gauge theory and resolving the spectrum of local
  operators,''
  Phys.\ Rev.\  D {\bf 78}, 126003 (2008)
  [arXiv:0807.3696 [hep-th]].

\bibitem{Das:2000st}
  S.~R.~Das, A.~Jevicki and S.~D.~Mathur,
  ``Vibration modes of giant gravitons,''
  Phys.\ Rev.\  D {\bf 63}, 024013 (2001)
  [arXiv:hep-th/0009019].

\bibitem{Sadri:2003mx}
  D.~Sadri and M.~M.~Sheikh-Jabbari,
  ``Giant hedge-hogs: Spikes on giant gravitons,''
  Nucl.\ Phys.\  B {\bf 687}, 161 (2004)
  [arXiv:hep-th/0312155].

\bibitem{Berenstein:2003ah}
  D.~Berenstein,
  ``Shape and holography: Studies of dual operators to giant gravitons,''
  Nucl.\ Phys.\  B {\bf 675}, 179 (2003)
  [arXiv:hep-th/0306090].

\bibitem{Balasubramanian:2002sa}
  V.~Balasubramanian, M.~x.~Huang, T.~S.~Levi and A.~Naqvi,
  ``Open strings from N = 4 super Yang-Mills,''
  JHEP {\bf 0208}, 037 (2002)
  [arXiv:hep-th/0204196],\\
  O. Aharony, Y.E. Antebi, M. Berkooz and R. Fishman, ``Holey sheets: Pfaffians and 
  subdeterminants as D-brane operators in large $N$ gauge theories,''
  JHEP {\bf 0212}, 096 (2002)
  [arXiv:hep-th/0211152].

\bibitem{Berenstein:2006qk}
  D.~Berenstein, D.~H.~Correa and S.~E.~Vazquez,
  ``A study of open strings ending on giant gravitons, spin chains and
  integrability,''
  [arXiv:hep-th/0604123],\\
  D.~Berenstein and S.~E.~Vazquez,
  ``Integrable open spin chains from giant gravitons,''
  JHEP {\bf 0506}, 059 (2005)
  [arXiv:hep-th/0501078],\\
  D.~Berenstein, D.~H.~Correa and S.~E.~Vazquez,
  ``Quantizing open spin chains with variable length: An example from giant
  gravitons,''
  Phys.\ Rev.\ Lett.\  {\bf 95}, 191601 (2005)
  [arXiv:hep-th/0502172],\\
  R.~de Mello Koch, N.~Ives, J.~Smolic and M.~Smolic,
  ``Unstable giants,''
  Phys.\ Rev.\ D {\bf 73}, 064007 (2006)
  [arXiv:hep-th/0509007],\\
  A.~Agarwal,
  ``Open spin chains in super Yang-Mills at higher loops: Some potential
  problems with integrability,''
  [arXiv:hep-th/0603067],\\
  K.~Okamura and K.~Yoshida,
  ``Higher loop Bethe ansatz for open spin-chains in AdS/CFT,''
  [arXiv:hep-th/0604100],\\
  D.~H.~Correa and G.~A.~Silva,
  ``Dilatation operator and the Super Yang-Mills duals of open strings on AdS
  Giant Gravitons,''
  [arXiv:hep-th/0608128].
  M.~Ali-Akbari and M.~M.~Sheikh-Jabbari,
  ``Electrified BPS Giants: BPS configurations on Giant Gravitons with Static
  Electric Field,''
  JHEP {\bf 0710}, 043 (2007)
  [arXiv:0708.2058 [hep-th]],\\
  V.~Balasubramanian, J.~de Boer, V.~Jejjala and J.~Simon,
  ``Entropy of near-extremal black holes in AdS$_5$,''
  JHEP {\bf 0805}, 067 (2008)
  [arXiv:0707.3601 [hep-th]].
  R.~Fareghbal, C.~N.~Gowdigere, A.~E.~Mosaffa and M.~M.~Sheikh-Jabbari,
  ``Nearing Extremal Intersecting Giants and New Decoupled Sectors in N = 4
  SYM,''
  JHEP {\bf 0808}, 070 (2008)
  [arXiv:0801.4457 [hep-th]],\\
  D.~Berenstein and D.~Trancanelli,
  ``Three-dimensional N=6 SCFT's and their membrane dynamics,''
  Phys.\ Rev.\  D {\bf 78}, 106009 (2008)
  [arXiv:0808.2503 [hep-th]],\\
  D.~M.~Hofman and J.~M.~Maldacena,
  ``Reflecting magnons,''
  JHEP {\bf 0711}, 063 (2007)
  [arXiv:0708.2272 [hep-th]],\\
  R.~I.~Nepomechie,
  ``Bethe ansatz equations for open spin chains from giant gravitons,''
  JHEP {\bf 0905}, 100 (2009)
  [arXiv:0903.1646 [hep-th]].

\bibitem{Kimura:2007wy}
  Y.~Kimura and S.~Ramgoolam,
  ``Branes, Anti-Branes and Brauer Algebras in Gauge-Gravity duality,''
  arXiv:0709.2158 [hep-th].

\bibitem{Brown:2007xh}
  T.~W.~Brown, P.~J.~Heslop and S.~Ramgoolam,
  ``Diagonal multi-matrix correlators and BPS operators in N=4 SYM,''
  arXiv:0711.0176 [hep-th].

\bibitem{Brown:2008rr}
  T.~W.~Brown, P.~J.~Heslop and S.~Ramgoolam,
  ``Diagonal free field matrix correlators, global symmetries and giant
  gravitons,''
  arXiv:0806.1911 [hep-th].

\bibitem{Kimura:2009wy}
 Y.~Kimura,
  ``Non-holomorphic multi-matrix gauge invariant operators based on Brauer
  algebra,''
  arXiv:0910.2170 [hep-th].
 
\bibitem{Brown:2008rs}
  T.~W.~Brown,
  ``Permutations and the Loop,''
  arXiv:0801.2094 [hep-th].

\bibitem{Ramgoolam:2008yr}
  S.~Ramgoolam,
  ``Schur-Weyl duality as an instrument of Gauge-String duality,''
  AIP Conf.\ Proc.\  {\bf 1031}, 255 (2008)
  [arXiv:0804.2764 [hep-th]].

\bibitem{tomyusuke}
T.~W.~Brown,
  ``Cut-and-join operators and N=4 super Yang-Mills,''
  arXiv:1002.2099 [hep-th],\\
 Y.~Kimura,
  ``Quarter BPS classified by Brauer algebra,''
  arXiv:1002.2424 [hep-th].

\bibitem{Beisert:2003tq}
  N.~Beisert, C.~Kristjansen and M.~Staudacher,
  ``The dilatation operator of N = 4 super Yang-Mills theory,''
  Nucl.\ Phys.\  B {\bf 664}, 131 (2003)
  [arXiv:hep-th/0303060].

\bibitem{de Mello Koch:2004ws}
  R.~de Mello Koch and R.~Gwyn,
  ``Giant graviton correlators from dual SU(N) super Yang-Mills theory,''
  JHEP {\bf 0411}, 081 (2004)
  [arXiv:hep-th/0410236].

\bibitem{Bhattacharyya:2008rc}
  R.~Bhattacharyya, R.~de Mello Koch and M.~Stephanou,
  ``Exact Multi-Restricted Schur Polynomial Correlators,''
  arXiv:0805.3025 [hep-th].

\end{thebibliography}
\end{document}